# Robust Evaluation of Neural Encoding Models
# via ground-truth approximation

Giovanni M. Di Liberto[1,2,3]

[1] School of Computer Science and Statistics, University of Dublin, Trinity College, Ireland; ADAPT Centre
[2] Trinity College Institute of Neuroscience, Trinity College Dublin, Dublin, Ireland.
[3] Trinity Centre for Biomedical Engineering, Ireland

Correspondence: gdiliber@tcd.ie

Conflicts of interest: none declared.

**Funding sources:** This research was supported by Research Ireland at ADAPT, the Research Ireland Centre for AI-Driven Digital Content Technology at Trinity College Dublin and University College Dublin [13/RC/2106_P2]. For the purpose of Open Access, the author has applied a CC BY-NC-ND public copyright licence to any Author Accepted Manuscript version arising from this submission.

**Acknowledgements:** I thank Prof. Edmund C. Lalor for useful feedback on the results and for suggesting the use of the match-vs-mismatch validation metric; Dr. Simon Geirnaert for excellent discussions on this metric and alternative approaches; I thank Mrs. Claire Donohue, Dr. Aaron Nidiffer, and the members of my research team at Trinity College Dublin for useful scientific discussions around the ideas at the core of this manuscript. I thank Yannick Peters for his work on the Open Data explorer and the list of datasets on the website https://diliberg.net. I thank the colleagues at the CNSP open-science initiative (https://cnspworkshop.net) for contributing to some of the resources used in this study, especially Jessica M. Alexander for the conversion of the GwilliamsSpeechMEG datasets to CND. Thanks to all the colleagues that have agreed to share their datasets publicly, contributing to this study and much more research in the future (please find a full list in **Table 1**).

Word count abstract: 150.

Word count (excluding abstract, title page, references and methods): 5059.

**Abstract**

Encoding models enable measurement of how our brains represent sensory inputs using electro- and magneto-encephalography (MEEG). Evaluating how closely encoding models reflect the underlying brain functions is a crucial premise for model interpretation and hypothesis testing. However, the ground-truth neural activity is unknown, preventing model evaluation with respect to the target neural signal. Existing evaluation metrics must therefore relate model's predictions to noisy MEEG measurements, where most variance is stimulus-unrelated. Here, I introduce an evaluation framework where model predictions are compared to a ground-truth approximation, obtained by aligning MEEG signals with predictions using canonical correlation analysis and via participant averaging. The resulting metric (CPA-PA) yields single-participant evaluations outperforming conventional scores by ~300-1000% on synthetic EEG data and ~250% on 34 real MEEG datasets (818 datapoints). These gains reflect increased sensitivity to stimulus-relevant neural activity and reduced dependence on SNR, establishing ground-truth approximation as a robust framework for evaluating encoding models.

## Main

Encoding models have become central to studying how the human brain represents sensory information using electro- and magneto-encephalography (MEEG). By predicting neural responses from stimulus features, these models allow quantification of neural encoding strength, representational structure, and computational transformations from sensory inputs to higher-level abstractions[1-6]. However, evaluating these models remains challenging due to the substantial noise of MEEG signals, where only a small fraction of sensor-level variance reflects stimulus-driven activity[7]. As a result, standard evaluation metrics, typically Pearson's or Spearman's correlation between predicted and recorded time-series, underestimating model performance. Improved evaluation metrics could lead to a clearer understanding of brain functions, especially when considering that MEEG recordings can be particularly noisy on certain participant cohorts (e.g., infants), tasks (e.g., involving movement), or technology (e.g., low-cost systems with low SNR). This study takes a step in that direction by introducing an evaluation framework for encoding models that informs on the model quality looking past the MEEG noise.

Sensory neuroscience widely adopts linear time-invariant (LTI) encoding models, as they offer a numerically stable, interpretable, and computationally efficient framework for linking multivariate stimuli to neural responses. The relative simplicity of these models, however, has also been blamed for their limited ability to explain MEEG variance. Here, I propose that LTI encoding models may have been systematically undervalued in sensory neuroscience research due to the fundamental limitations of conventional evaluation metrics, underestimating the extent to which these models capture sensory-relevant neural activity. This study presents a framework that solves that issue by comparing MEEG predictions with an approximation of the ground-truth neural signal, informing on the actual model's explanatory power through a noise-robust evaluation. The ground-truth approximation is obtained through Canonical Prediction Alignment (**CPA**), which suppressed stimulus-irrelevant signals in the MEEG recording by aligning model predictions and neural data using a Canonical Correlation Analysis (CCA) mapping (**Fig. 1a**). The MEEG signal can be further denoised via Participant Averaging (PA) when stimuli are shared across participants (**Fig. 1b**). Encoding models fitted on the original data are then evaluated against the resulting ground-truth approximation (**CPA-PA**).

The effectiveness of CPA and CPA-PA is explored using the LTI encoding model known as the Temporal Response Function (TRF), due to its wide use in continuous sensory experiments across sensory modalities[2,8-15]. Note that CPA can be applied to any encoding model[16,17], as the procedure aligns MEEG signals to their prediction, leaving the model itself untouched. The validation of CPA and CPA-PA is carried out systematically on both synthetic and real MEEG data, considering 800 synthetic participants in total, and 34 MEEG datasets, with 818 datapoints overall. Part of the validation is carried out on synthetic data, where the ground-truth signal is known and there is full control over the level of noise. Ground-truth signals were generated as the convolution of an artificially-built impulse response with the sound envelope from a widely studied audiobook-listening EEG dataset (*LalorNatSpeech*[18]). A realistic TRF waveform was obtained by modelling the impulse response as a sum of Gaussian components with timing and amplitude corresponding to the P1–N1–P2 matching empirical measurements[19]. Noise consisting of randomly selected EEG chunks was added to the resulting ground-truth signal. Real datasets were standardised and processed in a consistent manner, combining publicly available dataset with new unpublished data.

This study provides three main contributions. First, it demonstrates the effectiveness of evaluating encoding models via ground-truth approximation, introducing the CPA and CPA-PA metrics. Second, it reveals that conventional evaluation metrics severely underestimate the model's predictive power, with CPA-PA enhancing MEEG prediction correlation scores by ~300-1000% and ~250% on synthetic and real neural recordings respectively (**Fig. 1c,f**). These gains are more than a simple rescaling of correlation scores, as comparable increases were also measured for signal detectability (ability to classify matching versus mismatching stimulus-MEEG segments; **Fig. 1d,g**). Additional analyses are then carried out at the single-participant level, revealing that CPA can retrieve neural encodings where conventional metrics fail (**Fig. 2d-right**). These results are found to be consistent across 34 datasets (31 EEG, 2 MEG, 1 fNIRS; **Fig. 1e; Supplementary Fig. 1**), all pre-processed and analysed in the same manner. This brings us to the third contribution: A new publicly available resource including standardised datasets and code (https://osf.io/c76p8/overview[i]), including both previously published and new datasets, with tasks such as speech listening and production, music listening and imagery, video and sign language watching, selective auditory attention, and involving adult, children, and infant participants (**Table 1** and **Supplementary Table 1**).

# Results

## Validation on synthetic data

First, the effectiveness of the CPA and CPA-PA evaluation metrics is tested on univariate encoding models applied to SyntheticEEG-1, comprising six datasets for different SNR levels, each with 30 simulated participants. The stimulus-EEG mapping is studied by considering a single stimulus feature, the speech envelope, which was selected due to its wide adoption[19-22] and its availability across all the considered datasets[ii]. For any given participant, the CPA and CPA-PA correlation scores (R-CPA and R-CPA-PA) are compared with a conventional EEG prediction correlation score, where a single score is derived by averaging prediction correlations across all MEEG channels (R-AVG). Results are also reported for the channel with highest prediction correlation (R-MAX), where the optimal channel is determined with cross validation, testing if R-CPA is superior to the prediction correlation at any single channel. All evaluation scores were calculated on held-out folds that were not included on the TRF and CCA model fit (see **Online Methods**).

As hypothesised, CPA evaluation was more sensitive to the neural ground-truth signal than the conventional evaluation metrics, leading to EEG prediction correlation gains for R-CPA over R-MAX of 4%, 24%, 127%, 331%, 506%, and 1012%, for SNRs 0, -10, -20, -30, -40, and -50 dB respectively, calculated as *(R-CPA - R-MAX)/R-MAX* (**Fig. 1c**). Even larger gains emerged for R-CPA-PA over R-MAX of 4%, 27%, 150%, 516%, 938%, 3241%. The result was confirmed by a two-way repeated-measures ANOVA with factors *Metric* (R-AVG, R-MAX, R-CPA, and R-CPA-PA) and *SNR* (6 levels). The analysis revealed robust main effects of *Metric* ($F(3, 87) = 6023.6$, $p = 9.4 \times 10^{-101}$), *SNR* ($F(5, 145) = 696.8$, $p = 1.8 \times 10^{-99}$), and a *Metric × SNR* interaction ($F(15, 435) = 1374.3$, $p < 10^{-101}$). *Post hoc* tests confirmed that R-CPA-PA > R-CPA (FDR-corrected paired two-tailed Wilcoxon signed-rank test, $p < 10^{-3}$ for all SNRs), and R-CPA > R-MAX for all SNRs ($p < 10^{-5}$), except SNR 50 ($p = 0.063$; exact p-values and effect sizes are reported in the supplementary data).

---

[i] see also Open Data section in https://diliberg.net

[ii] A visual equivalent of the sound envelope, the instantaneous visual change, was used on the Sign Language datasets

To verify that these gains reflect improved sensitivity to the task-relevant neural signal rather than a simple numerical rescaling, we quantified *signal detectability* using a match-vs-mismatch framework[23-25]. This framework is grounded in the intuition that a good encoding model should generate predictions that more closely resemble the neural response corresponding to the correct stimulus segment ("match") than the responses associated with a randomly selected stimulus segment ("mismatch"). Signal detectability calculated over 5-second segments (**Fig. 1d**) confirms the trends measured for EEG prediction correlation metrics. Specifically, numerical results indicate EEG prediction correlation gains for R-CPA over R-MAX of 0%, 0%, 1%, 43%, 257%, 3550% for SNRs 0, -10, -20, -30, -40, and -50 dB respectively (**Fig. 1d**), with larger gains when comparing R-CPA-PA with R-MAX of 0%, 0%, 1%, 44%, 343%, 9235%. A two-way repeated-measures ANOVA revealed robust main effects of *Metric* ($F(3, 87) = 1458.6$, $p = 3.0 \times 10^{-74}$), *SNR* ($F(5, 145) = 234.96$, $p = 1.2 \times 10^{-67}$), and a *Metric × SNR* interaction ($F(15, 435) = 428.0$, $p < 10^{-249}$). *Post hoc* tests confirmed that R-CPA-PA > R-CPA for SNRs -30dB (FDR-corrected paired two-tailed Wilcoxon signed-rank test, $p = 2.2 \times 10^{-4}$), -40dB ($p = 1.7 \times 10^{-6}$), and -50dB ($p = 2.0 \times 10^{-4}$), while no statistically significant effect was measured for SNRs where signal detectability was saturating close to 100% accuracy i.e., 0dB ($p = 1$), -10dB ($p = 1$), -20dB ($p = 0.063$). *Post hoc* tests also indicated that R-CPA > R-MAX for all SNRs more difficult than -10dB ($p < 10^{-3}$ for all comparisons), apart from SNRs 0dB ($p = 1$) and -10dB ($p = 1$), where accuracy was close to 100% for both metrics.

## Real-data validation: Encoding models achieve substantially higher signal detectability than previously recognised

Univariate encoding models were run on 34 real MEEG datasets (**Fig. 1e**; **Table 1**) demonstrating that, for a given encoding model, R-CPA and R-CPA-PA evaluation metrics produce larger correlation scores (**Fig. 1f**) and signal detectability (**Fig. 1g**) than the conventional metrics R-AVG and R-MAX. An average MEEG prediction correlation gain of 68% was measured for R-CPA over R-MAX, with an even larger gain of 252% when comparing R-CPA-PA over R-MAX. The result was confirmed by a one-way repeated-measures ANOVA with the factor *Metric* (R-AVG, R-MAX, R-CPA) and 34 datasets, revealing robust main effects of *Metric* ($F(2, 66) = 61.1$, $p = 9.6 \times 10^{-16}$). *Post hoc* tests confirmed that R-CPA > R-MAX (FDR-corrected paired two-tailed Wilcoxon signed-rank test, $p = 4.0 \times 10^{-7}$; effect-size-r = 0.87). A separate repeated-measures ANOVA was run for comparing R-CPA-PA with the other metrics, as that metric could only be calculated on a subset of data where participants were presented with the same stimuli (21 datasets). A statistically significant effect of *Metric* (R-AVG, R-MAX, R-CPA, R-CPA-PA) was detected ($F(3, 60) = 59.8$, $p = 5.1 \times 10^{-18}$). *Post hoc* tests confirmed that R-CPA-PA > R-CPA ($p = 6.9 \times 10^{-5}$; effect-size-r = 0.88).

Consistent with the correlation analysis, R-CPA and R-CPA-PA showed substantial improvements in *Signal Detectability* over conventional metrics (**Fig. 1g**), with gains over R-MAX of 71% and 210% for R-CPA and R-CPA-PA respectively, confirming the trends seen for the prediction correlations. These gains reflected a statistically significant effect of *Metric* (R-AVG, R-MAX, R-CPA) on the 34 datasets ($F(2, 66) = 73.2$, $p = 1.8 \times 10^{-17}$). *Post hoc* tests confirmed that R-CPA > R-MAX (FDR-corrected paired two-tailed Wilcoxon signed-rank test, $p = 9.6 \times 10^{-7}$; effect-size-r = 0.86). A statistically significant effect of *Metric* (R-AVG, R-MAX, R-CPA, R-CPA-PA) was also detected when considering R-CPA-PA on the corresponding 21 datasets ($F(3, 60) = 1520.9$, $p = 1.8 \times 10^{-25}$). *Post hoc* tests confirmed that R-CPA-PA > R-CPA ($p = 1.2 \times 10^{-4}$; *effect-size-r* = 0.88).

## Retrieving neural signatures at the single-participant level

To test whether CPA enhances the recovery of neural variables of interest at the single-participant level, a second synthetic dataset was constructed (SyntheticEEG-2; N=50), designed to dissociate SNR-driven variability from meaningful differences in neural encoding. Unlike the previous simulation, individuals within each SNR group had variable SNR values (mean ± 2.5 dB, uniformly sampled). This makes it possible to quantify the sensitivity of each evaluation metric to difference in SNR between participants. The simulated EEG signal consisted of a combination of neural responses to the speech envelope and to word onsets, generated from spatially distinct neural sources (**Fig. 2a**). Crucially, the strength of the word-onset response was modulated by a scaling factor $\beta$. For the sake of a simple intuition, $\beta$ can be regarded as a proxy of individual language proficiency, such that higher proficiency corresponds to a stronger neural response to word onsets. This design enables a direct test of whether a given evaluation metric is primarily sensitive to SNR variability, or whether it can reliably capture meaningful differences in neural signals associated with $\beta$.

Multivariate forward TRFs were fit to capture the mapping from speech envelope and word onsets to the synthetic EEG data. Visual inspection indicates that averaging the TRF weights across all 50 participants leads to an accurate retrieval of the envelope and word onset temporal response functions that were injected in the synthetic data (**Fig. 2b**). When reducing the number of participants to 10, the envelope TRF can still be retrieved at all SNR levels, while a more difficult and imprecise retrieval emerges at the more challenging SNR levels (-40 and -50 dB). This qualitative observation is quantified by measuring the correlation between the target variable $\beta$ and the TRF value for lag 400ms at Pz, where the word onset component was artificially centred. Strong correlations were detected for SNRs 0 dB ($r$ = -0.76, $p$ = $10^{-10}$), -10 dB ($r$ = -0.74, $p$ = $8\times10^{-10}$), -20 dB ($r$ = -0.79, $p$ = $6\times10^{-12}$), and -30 dB ($r$ = -0.69 $p$ = $3\times10^{-8}$), with significant yet decreasing correlations at the more challenging SNRs -40 dB ($r$ = -0.46 $p$ = $9\times10^{-4}$) and -40 dB ($r$ = -0.28 $p$ = 0.05). This indicates that the encoding models remain sensitive to the neural variable of interest $\beta$ despite the high levels of noise. While that ability to retrieve neural responses of interest is encouraging, do conventional evaluation metrics exploit that ability in full or are they a limiting factor in what TRFs can do?

## Discerning sensory-related neural sources with CPA

Multivariate TRFs derived for SyntheticEEG-2 were evaluated with R-AVG, R-MAX, R-CPA, and R-CPA-PA. The numerical results indicate that CPA and CPA-PA exhibit greater sensitivity to sensory-related neural signals, allowing model explanatory power to be detected across all SNR levels, even in conditions where conventional metrics fail to reveal any relationship (see SNR=−50dB in **Supplementary Fig. 3**). This increased sensitivity provides a more reliable criterion for determining whether an encoding model can be interpreted neurophysiologically or not, thereby extending the applicability of encoding models such as forward TRFs to more challenging neural recordings.

In addition to serving as that critical threshold, evaluation metrics have also been proven useful approaches for determining if neural signals encode a certain stimulus feature of interest[5,19,24] and its encoding strength (e.g., envelope tracking)[26-31]. An additional analysis was carried out on SyntheticEEG-2 to determine if the parameter $\beta$, which directly scales to word-onset TRF weights, could also be recovered using forward-model evaluation scores. Multivariate TRFs fit using both speech envelope and word onset were evaluated by predicting EEG in held-out data using word onset alone. Performance was quantified using EEG prediction correlation at the best channel (R-MAX-PARTIAL) and compared with R-CPA computed on the first two canonical components (R-CPA$_{comp1}$ and R-CPA$_{comp2}$), under the hypothesis that the two simulated neural sources would be disentangled across canonical components. As expected, R-CPA (i.e., R-CPA$_{comp1}$) and R-CPA$_{comp2}$ exhibited spatial patterns consistent with envelope-related and word-onset-related TRFs, respectively (**Fig. 2d**). Bothg R-MAX$_{partial-x2}$ and R-CPA$_{comp2}$ showed strong correlations with β (up to $r \approx 0.8$), but with a critical distinction: R-MAX$_{partial-x2}$ performed best at high SNRs (0 and -10 dB), whereas R-CPA$_{comp2}$ consistently outperformed all other metrics at more realistic low-SNR levels (-20 to -50 dB; **Fig. 2d**). With regard to sensitivity to SNR, R-MAX showed a large dependency on SNR fluctuations, with correlations up to $r \approx 0.7$. Instead, R-CPA had typically either comparable or lower values, never going past $r \approx 0.4$ (**Fig. 2d**).

A similar analysis was carried out on the *LalorNatSpeech* (audio-book listening task) and *LalorRevSpeech* (time-reverse speech listening) datasets, expecting the former to exhibit neural signatures of both acoustic (envelope) and lexical (word-onset) processing and the latter to encode acoustic but not lexical processing. As expected and in line with previous work[4,19], when considering R-MAX, a mixed-effects ANOVA revealed a statistically significant main effect of feature ($F(1, 54) = 9.6$, $p = 3.5×10^{-14}$) and a feature × group interaction ($F(1, 54) = 4.6$, $p = 0.036$), with no main effect of group ($F(1,54) = 0.5$, $p = 0.467$). *Post hoc* analyses confirmed a statistically significant between-group difference in lexical encoding (Wilcoxon rank-sum test, $p = 0.0124$; *effect-size-r* = 0.46), with no statistically significant group difference for acoustic features ($p = 0.2240$; *effect-size-r* = -0.23). When considering R-CPA, a statistically significant main effects of component ($F(1, 54) = 47.7$, $p = 5.8×10^{-9}$) and the between-group effect approaching statistical significant ($F(1, 54) = 3.2$, $p = 0.079$), with no feature × group interaction ($F(1, 54) = 0.02$, $p = 0.865$). *Post hoc* tests confirm a statistically significant group difference in CC2 (lexical spatial pattern) ($p = 0.0261$; *effect-size-r* = 0.41), with no statistically significant group difference emerged for CC1 (acoustic spatial pattern) ($p = 0.2240$; *effect-size-r* = 0.23).

Relating CPA with direct CCA modelling

CPA is built via a CCA mapping between the model predictions and the actual neural signal, which is then tested on held-out data. When the encoding model is a forward TRF, CPA can be seen as a concatenation of a TRF transformation *W*, a CCA forward model *A*, and a CCA backward model *B*. Since all these transformations are linear, this can be mathematically equivalent to a direct CCA mapping under certain conditions (see **Discussion**). There are some caveats in that regard, as both TRFs and CCA in this study include Ridge regularisation. Nonetheless, it is intuitive to understand that a CCA mapping, in particular the CCA-1 (lags are applied to the stimulus) mapping as defined by de Cheveigné and colleagues[32], should lead to correlation scores close or identical to R-CPA. This comparison was carried out on SyntheticEEG-2 (-30 dB) and *LalorNatSpeech*, in both cases revealing very similar correlations scores for R-CPA and CCA-1 across all components, who were outperformed by CCA-2 across all canonical components (lags applied to stimulus and EEG). Interestingly, the highest correlation was achieved by R-CPA-PA, whose first canonical component outperformed all other models (**Fig. 2f**).

The advantage of CPA over CCA mappings lies primarily in its conceptual and methodological separation between model estimation and model evaluation. CPA enables a more accurate evaluation of encoding models like the TRF, without altering the model. This means combining the good qualities of TRFs (such as its interpretability and stability) and CCA (such as its strong denoising quality), with little risks. In contrast, direct CCA approaches yield high correlation scores at the expense of interpretability, as stimulus-related information is distributed across a larger number of components, particularly in more complex mappings such as CCA-2 (**Fig. 2f**). CPA and CCA-1, instead, capture most of the stimulus-related neural signal with two components with the synthetic data, and three-four components on the real data.

An additional drawback of direct CCA mappings is their increased susceptibility to overfitting. In CPA, the CCA transformation operates without time-lag expansion and within the same sensor space, whereas CCA-1 and CCA-2 directly map stimulus features to neural data through high-dimensional lagged representations, including features of heterogeneous type. This increased model flexibility can substantially elevate overfitting risk relative to the more constrained TRF + CCA architecture used in CPA. Consistent with this hypothesis, CPA and CCA-1 were compared across 16 MEEG datasets involving continuous speech listening. Overfitting was quantified as the difference between training and test correlation for the first canonical component. CCA-1 exhibited significantly greater overfitting (M = 0.063, SD = 0.024) than CPA (M = 0.044, SD = 0.027; *t*-test, *p* = 0.014).

## Discussion

This study introduces a ground-truth approximation framework for the robust evaluation of encoding models. Compared with conventional evaluation metrics, this framework represents a substantial methodological advance, yielding evaluation scores that are orders of magnitude higher (**Fig. 1c,f**) and demonstrating enhanced sensitivity to task-related neural signals (**Fig. 1d,e**; **Fig. 2d**). Central to this framework is the CPA metric, where model performance is assessed in a denoised canonical space and thereby offers substantially increased sensitivity to stimulus-relevant neural activity. CPA-PA further enhances signal detectability in experimental designs involving shared stimuli by leveraging cross-participant consistency, at the cost of introducing assumptions about group homogeneity. Together, these metrics establish a general, add-on evaluation strategy that can be seamlessly integrated into existing encoding pipelines.

### The benefit of increased signal detectability

The ground-truth approximation evaluation indicate that LTI encoding models reported in the literature are far more effective at capturing sensory-related neural signals than previously suggested (**Fig. 1g**; **Supplementary Fig. 1 and 2**). Importantly, this improvement does not arise from changes to the encoding models themselves, but from a re-evaluation of the same model predictions using a more sensitive metric. As such, encoding models from earlier studies can be re-analysed using CPA or CPA-PA without re-estimating model weights. But what are the specific reasons making this increased signal detectability desirable?

Model evaluation is crucial for interpreting encoding models, as meaningful analysis of model weights is typically restricted to cases where statistical significance is reached. Model weights obtained from fits that do not exceed the significance threshold should not be trusted instead, as there is no guarantee that they reflect explainable, sensory-related neural activity[1]. However, SNR is a primary factor underlying the low evaluation scores produced by conventional metrics such as R-AVG and R-MAX (**Fig. 1c,d**; **Fig. 2c,d**). The consequence is that these metrics can fail to detect explainable sensory-related neural activity in low-SNR data, even when a model is, in fact, capturing valuable sensory neural activity (as observed for the SNR=-50dB condition in SyntheticEEG-2; **Fig. 2c** and **Supplementary Fig. 3**).

By increasing signal detectability, the proposed framework unlocks the use of encoding models in challenging datasets that were previously inaccessible, including datasets with low SNR or short recording durations due to the particular population (e.g., infants[13,33-35]), the task (e.g., interactive paradigms such as conversations[36-38]), and experimental settings (e.g., recordings collected "in-the-wild"[39]). One important consideration is that the most challenging datasets, where conventional evaluation metrics might have failed, are seldom publicly available. Yet these are precisely the conditions under which the strongest gain in signal detectability are likely to emerge, as illustrated by the analyses on the synthetic EEG data.

An important implication is that models previously considered insensitive to sensory-related information may in fact encode meaningful task-related signals, thereby justifying a closer examination of their learned weights and spatial-temporal structure. This was the case for low-SNR synthetic data, where R-AVG and R-MAX had chance-level (~50%) signal detectability, suggesting that the model fails to capture any sensory neural signal of interest (**Supplementary Figure 3**). On the same model, instead, CPA-PA indicates a much higher signal detectability (~74%), revealing that the model captures valuable sensory neural signal (which was the case by construction, as quantified in **Fig. 2c**). A similar pattern also emerged in real EEG recordings from infants, where CPA-PA revealed sensitivity to sensory signals where conventional metrics failed to detect reliable relationships (**Supplementary Fig. 1**)[34].

The proposed framework also increases the sensitivity to meaningful single-participant variability. As illustrated by the β-factor manipulation in **Fig. 2**, CPA preserves inter-individual differences more reliably than standard metrics, enabling finer-grained examination of participant-specific neural encoding properties. This is particularly relevant for developmental, clinical, or heterogeneous populations, where inter-individual variability is of primary scientific interest.

## Why re-evaluating the same model instead of using a "better" model altogether

Previous work has attempted to address the low sensitivity of encoding model evaluations by adopting alternative model mappings, including more elaborate regression frameworks[40], CCA[41], and deep learning (primarily focussing on envelope decoding however)[17,42]. However, a key motivation for the present approach is to optimise the evaluation metric of any given encoding model that one might intend to use. For example, TRFs have several desirable properties, including robustness, a small number of hyperparameters, rapid computation and, most importantly, ease of interpretability[2]. In many domains of sensory neurophysiology, interpretability is a non-negotiable requirement. The ground-truth approximation framework is therefore conceived as an add-on evaluation layer, rather than a replacement for established encoding models. This allows researchers to select a model structure that best serves their theoretical goals (e.g., TRFs for interpretability), while still benefiting from substantially improved evaluation sensitivity.

Alternative approaches like deep learning enables non-linear mappings at the cost of a more challenging model fit, parameter tuning, and interpretability. A successful workaround is instead to overcome linearity constraints by using LTI models after elaborate non-linear features extraction procedures (e.g., using LLMs)[19,43,44]. An alternative approach that can yield high correlation scores is CCA, which offers a conceptually elegant combination of encoding and decoding models, which are calculated simultaneously as a multivariate-to-multivariate mapping[23]. However, that comes at the cost of an increased model complexity compared to a simpler LTI encoding model, with a more challenging model interpretation and a higher risk of overfitting (see **Fig. 2f** and the Result section "*Relating CPA with direct CCA modelling*").

Intermediate solutions have been attempted such as back-to-back regression[40], which also combines encoding and decoding models, but as a chain. Forward models are evaluated by measuring how well MEEG predictions can reconstruct each stimulus feature, where the latter is estimated using decoding models. While this evaluation can be effective — because correlations are computed in stimulus space, where the ground-truth signal is known — this approach inherits key limitations of decoding models. First, evaluation scores are computed separately for each stimulus feature separately (a multivariate-to-univariate mapping), providing multiple scores instead of a single value. Second, the reconstruction scores are calculated for features that can be quite different in nature (e.g., continuous versus discrete), thus producing scores that may not be directly comparable. This is because reconstruction correlation depends, in addition to neural encoding strength, to the temporal structure, sparsity, and dynamic range of the stimulus feature itself, all of which differ substantially across feature type (e.g., sound envelope and word onsets).

CPA solves that issue by leaving the encoding model untouched, thus maintaining its original interpretability, and by carrying out the model evaluation in a denoised MEEG space approximating the neural ground-truth signal. In doing so, CPA combines the desired benefits of CCA and back-to-back regression, while circumventing their shortcomings. CPA-PA pushes even further by enhancing the ground-truth approximation, leading to substantial gains in signal detectability.

## Ground-truth approximation: Design choices and alternatives

The aim of evaluation via ground-truth approximation is to derive metrics that are more strongly correlated with the underlying task-related neural activity. The results presented here confirm that CPA achieves this goal across all scenarios considered. Nevertheless, it is important to reflect on the specific design choices for CPA and PA, and whether those influence how the results should be interpreted. CPA operates under the same assumptions as conventional evaluation of encoding models. Its advantage arises from evaluating predictions in a denoised canonical space, achieved via a direct CCA mapping, thus via spatial filtering (similar to the well-known independent component analysis, but where the spatial maps are derived to maximise stimulus-neural correlations). This entails the assumption that noise and task-related neural sources can be separated primarily along the spatial dimension.

In principle, the denoising could be further strengthened by incorporating temporal filtering or time-lagged representations[23]. However, this would further increase the risk of overfitting and lead to a broader informational spread across canonical components (e.g., CCA-2 in **Fig. 2f**), complicating interpretability. As such, the present study opted for a lag-free CCA mapping, yielding a solution that is effective, stable, interpretable, and computationally efficient.

The PA operation introduces a qualitatively different assumption, namely that neural activity is consistent across the participant cohort. While such an assumption is often part of ERP/ERF or TRF reporting, it directly affects single-participant evaluation by quantifying alignment with participant-averaged neural activity and attenuating sensitivities to participant-specific neural reactions. This becomes particularly relevant when comparing groups[31,45-47], as R-CPA-PA is influenced by within-group homogeneity, highlighting the importance of also reporting metrics insensitive to group homogeneity, such as R-CPA, or explicitly quantifying homogeneity using complementary measures such as inter-subject correlation[48]. A methodologically more elegant solution is multiway CCA[49], which identifies components that are maximally correlated across participants. A key distinction is that PA assumes a shared set of spatial maps across all participants, whereas multiway CCA allows participant-specific spatial maps. This makes multiway CCA a more flexible framework, albeit at the cost of increased computational complexity and a larger hyperparameter space. Future work may explore integrating multiway CCA into the ground-truth approximation procedure.

## Measuring neural tracking

In sensory neurophysiology, neural tracking is a central concept for understanding how continuous sensory streams are represented and processed in the brain. Over the last decade, input–output modelling approaches, including encoding and decoding models, have become widely used for measuring neural tracking[50]. Neural tracking varies across cortical sources and scalp locations, making forward models well suited to this task in principle. However, the low correlation values produced by conventional evaluations (**Fig. 1c**; **Supplementary Fig. 1,2**) and their strong dependence on SNR variability (e.g., **Fig. 2d**) have led many studies to rely instead on decoding models[29,50-54]. Neural tracking is then quantified as the ability to reconstruct a stimulus feature such as the speech envelope (i.e., envelope tracking) from distributed neural activity.

Model interpretation, however, cannot be carried out in the decoding direction. Many studies therefore combine decoding and encoding approaches, using decoding models to establish the presence of neural tracking and forward encoding models to interpret spatial or temporal weights[55-58]. This practice entails interpreting parameters of a model that was not used to quantify the underlying effect. Partial solutions, such as mathematically transforming decoding weights into an encoding space, have been proposed, but are often overlooked in favour of fitting separate models. That can be partly solved by deriving the equivalent encoding weights from the decoder mathematically[59].

A further limitation of decoding-based neural tracking is that it typically addresses one feature at a time, which is problematic for analyses involving multiple, mixed-type features such as spectrotemporal representations, phonetic features, or higher-level linguistic variables (e.g., lexical surprise or entropy). Forward encoding models naturally accommodate mixed-type feature spaces and enable comparisons within a unified neural space. By addressing the issue of low evaluation scores, the ground-truth approximation framework enables robust measurement of neural tracking using a single forward encoding model, thereby unifying estimation and interpretation within the same modelling framework.

## Beyond EEG and LTI models

The proposed ground-truth approximation methodology is inherently flexible, as it is only requires model predictions and the acquired neural signal for implementing CPA and CPA-PA. As such, it can be applied to any encoding model architecture and is not restricted to LTI models or to specific types of input features. While most demonstrations in this study focus on EEG, the framework is also validated on MEG and fNIRS data. That validation was carried out with configurations that are appropriate across all dataset considered, enabling their direct comparison (**Supplementary Fig. 1 and 2**). While some exceptions were made (for example, slower temporal dynamics were considered with fNIRS), future work should explore further technology specific configurations, such as higher frequency rates in MEG or source space analysis. Similarly, the majority of the data considered involved listening tasks, as that represents the majority of the publicly available data to date. To also explore the effectiveness of CPA and CPA-PA beyond listening paradigms, additional datasets are considered, with new datasets shared, with tasks including speech production, receptive sign language, comedy show watching, and music imagery[60,61].

## A new standardised resource for sensory neurophysiology

Another major contribution of this work is the release of a new standardised open resource comprising 34 MEEG and fNIRS datasets, together with all analysis code used in this study (https://osf.io/c76p8/overview, https://diliberg.net). By combining previously published and newly shared datasets spanning speech listening and production, music perception and imagery, sign language, video watching, and selective attention, including adult, child, and infant participants, this resource enables large-scale, reproducible comparisons of encoding models and evaluation metrics under uniform preprocessing and analysis choices.

# Online Methods

## Technical details of CPA and CPA-PA

Encoding models fit on a given subset of a dataset (fold 1) are conventionally evaluated by building MEEG signal predictions on left-out data (folds-2) and then measuring MEEG prediction correlations, which inform on the variance explained by the model. The evaluation framework presented here consists of comparing the model's predictions with a ground-truth neural signal, instead of the actual MEEG signal. This achieved by adding two additional steps: Canonical Prediction Alignment (CPA) and Participant Averaging (PA). Together, these methodologies produce a ground-truth approximation that, when compared with the MEEG predictions, informs on the model's ability to capture stimulus-relevant neural activity, diminishing the sensitivity to stimulus-irrelevant noise.

The first additional step is CPA: a CCA mapping[23] is fit on fold-2 to project MEEG signals and their predictions into a shared latent space where they are maximally aligned (**Fig. 1a**). Given a third fold unseen by both the encoding model and the CCA (fold-3), MEEG predictions are generated using the encoding model (previously fit on fold-1), and both the predicted and real MEEG signals are projected into the shared latent space using the CCA model (previously fit on fold-2). The CPA score is computed as the Pearson correlation at the first canonical component (CC), where stimulus-related activity is maximised.

Note that the description above uses the labels fold-1, fold-2, and fold-3 for conceptual clarity. In practice, these roles are achieved within a full nested cross validation loop, where the specific fold assignments differ at each iteration. This nested loop scheme optimises data usage and ensure stable model estimation. Nested cross-validation avoids the optimistic bias that arises when the same data are used for both hyperparameter selection and performance estimation[62,63]. In this case, hyperparameter selection involved both the encoding model and the CCA (find the list of hyperparameters in the respective subsections).

The second addition to the evaluation procedure is PA: MEEG signals from multiple participants is time-aligned and averaged on fold-2 and fold-3. The encoding model fit in fold-1 is evaluated as for CPA, with the difference that the model predictions in fold-2 and fold-3 are compared with MEEG data after participant averaging. PA was introduced (albeit without naming it) in a previous study as a solution for increasing signal detectability in EEG responses to nursery rhymes in infants and adults[34], building on previous research on a related approach[64] and inspired by methods attempting to isolate a ground-truth neural signal by combining data from multiple participants[49]. In practice, PA denoises the MEEG recording, reducing the influence of signals that are stimulus-irrelevant and inconsistent across participants. Together, CPA and PA inform on the encoding model's sensitivity to stimulus-relevant neural activity via a ground-truth approximation (**Fig. 1b**). One caveat must be noted for PA: it is only applicable in datasets where each given stimulus is presented to more than one participant and, ideally, where all participants are presented with the same stimuli (regardless of the presentation order). As such, this study presents results for both CPA, which is universally applicable, and CPA-PA, which applies to a subset of datasets (in this case, 21 out of 34 datasets).

**Inputs and outputs**

Inputs and outputs, typically stimulus features and neural responses respectively, are represented as multivariate time-series matrices X ($T \times d_1$) and Y ($T \times d_2$), where $T$ denotes the number of time samples. Inputs and outputs must be temporally aligned and (down)sampled at identical rates. The framework supports mixed feature types, including continuous and temporally-sparse. While the neural signal typically consists of continuous time-series, the possibility of using mixed data types gives important flexibility when selecting the stimulus features (e.g., speech envelope, word onsets). Note that most analyses in this study were carried out using the sound envelope feature, as that was available across all datasets (except the two sign language datasets, where the sound envelope was replaced by its visual equivalent – instantaneous visual change[65]).

**Encoding model – the Temporal Response Function**

The encoding model used here to analyse the mapping between stimulus features and MEEG signal is known as the temporal response functions (TRF) framework[9,15,66]. A TRF can be interpreted as a filter describing a linear time-invariant (LTI) transformation from X to Y, i.e.,

$$Y(t) = TRF * X(t),$$

where * represents the convolution operator. TRFs were estimated using lagged ridge regression, where the input $X_{lags} \in \mathbb{R}^{T \times (d1 \cdot K)}$, comprises the stimulus features $X \in \mathbb{R}^{T \times d1}$, including $K$ time-lagged versions of those features over a pre-determined time-lag window where the neural response is expected. In this study, the time-lag window was fixed at [-100, 600] ms (except with the *PodcastListening_fNIRS* dataset, where [-2,10] seconds was used to capture the slow dynamics of the hemodynamic response).

The TRF performs a multivariate-to-univariate mapping, meaning that a TRF is fit at each channel separately, returning model weights corresponding to each feature and time-lag. The weights $w \in \mathbb{R}^{d1 \times K}$ are derived to minimise the mean-squared error (MSE) between the predicted neural response and the measured EEG signal. Specifically, for the stimulus matrix $X_{lags}$ and the neural response vector y at a given channel in Y, the ridge estimator solves the minimisation problem:

$$\hat{w} = \text{argmin}_w \; (||\, y - X_{lags}\, w\, ||^2 + \lambda\, ||\, w\, ||^2),$$

where $\lambda$ is the L2 regularisation parameter and $||\, y - X\, w\, ||^2$ is the squared prediction error, while $||w||^2$ penalises large coefficients. The identity matrix is defined as $I \in \mathbb{R}^{(d1 \cdot K) \times (d1 \cdot K)}$. Here, the optimisation problem is solved with the closed-form formula:

$$\hat{w} = (X^T X + \lambda\, I)^{-1} X^T y$$

which yields the set of TRF weights across all time-lags and stimulus features for each channel. The regularisation parameter of the ridge regression ($\lambda$) is determined via cross validation, exploring the range of values from $10^{-2}$ to $10^6$ with logarithmic steps.

TRFs offer two views into the input-output relationship. The first view involves evaluating the strength of the X-Y mapping, which is typically quantified as the correlation between Y and its prediction on left-out folds with leave-one-out cross-validation. Here, MEEG prediction correlations were obtained using the Pearson's correlation score, returning one R-value for each MEEG channel. The score **R-AVG** is the average MEEG prediction correlation value across all channels. In this study, I also define the **R-MAX** score, which is the MEEG prediction correlation at the best-performing channel. To control for overfitting, the optimal channel index was selected within the inner loop of the nested scheme, and evaluated on held-out data in the outer loop.

**Canonical Prediction–MEEG Alignment (CPA)**

CPA evaluates encoding models by aligning predicted and recorded neural responses in a shared latent space derived through CCA. The CPA algorithm is implemented as follows:

a) For each cross validation step in the outer loop, one MEEG segment is held out for evaluation (fold-3).
b) For each step in the inner loop, one MEEG segment is held out (fold-2). The TRF models is fit on the remaining MEEG data (fold-1) and then used to build predictions on fold-2. The optimal value of λ is selected based on R-AVG in fold-1. This process is repeated for another selection of folds 1 and 2, until MEEG predictions are built for all MEEG segments in the inner loop.
c) After completion of the inner loop for the TRF model fit and MEEG prediction, a ridge regularised CCA mapping is fit between the TRF-derived MEEG predictions and the actual MEEG recordings. Before doing that, the optimal regularisation parameter ($λ_{CCA}$) is determined via a second cross validation inner loop. The search was constrained to the small set [1,10,100] for computational reasons. Note that, differently from the TRF and conventional CCA methods for stimulus-EEG mappings[32], the CCA mapping here did not include any time-shifts, as its purpose is to spatially align signals that are already time-aligned and in the same domain.
d) Correlations are calculated for each CC. In univariate models (using only one stimulus feature), R-CPA is the correlation at the first CC, where the stimulus-related variance is strongest. In multivariate models, multiple spatial maps are possible. As such, this study also reports correlations at multiple CCs (e.g., R-CPA-CC1, R-CPA-CC2).

**CPA-PA: Ground-truth approximation via participant averaging**

CPA-PA extends CPA by replacing fold-3 single-participant neural data with an averaged neural time-series obtained across participants, providing a high-SNR approximation of the stimulus-driven neural response. This way, the same models as before (fit on single-participant data) can be evaluated on a ground-truth approximation achieved by both single-participant CCA and group-averaging producing a single-participant CPA-PA score. CCA dimensionality was kept full-rank. Future work may explore dimensionality reduction to optimise efficiency.

**Signal detectability metrics**

The concept of signal detectability is based on the intuition that a good encoding model should produce prediction that more closely resemble the neural response recorded for the *correct* stimulus segment ("match") than for unrelated segments ("mismatch"). These match-vs-mismatch comparisons were performed for each of the four evaluation metrics considered i.e., R-AVG, R-MAX, R-CPA, and R-CPA-PA, after partitioning the MEEG data into 5-second long segments (except for the BabyRhythmCambridge datasets, where a window of 3-seconds was used to ensure a sufficient number of segments, due to the short duration of certain experimental trials). The resulting match-vs-mismatch classification accuracy is referred to as *signal detectability*, where 50% is the ideal chance level and the maximum score is 100%. A high signal detectability score reflects the model's ability to detect stimulus-relevant neural structure against the background noise. Signal detectability gains were calculated as:

$$\frac{SignalDetectability_{R-CPA} - SignalDetectability_{R-MAX}}{SignalDetectability_{R-MAX} - idealChanceLevel}.$$

## Realistic Synthetic data

### SyntheticEEG-1

Synthetic EEG datasets were built to test the ability of the encoding models and the evaluation metrics to retrieve stimulus-relevant information from neural recordings. Data was generated for each six different signal-to-noise rations (SNR) to determine how model performance are affected by the noise level, exploring SNRs of 0, −10, −20, −30, −40, and −50 dB. Thirty simulated participants were built for each SNR level. Synthetic EEG ground-truth signals ($y_1$) were generated as the convolution of the sound envelope from a widely studied audiobook-listening EEG dataset (*LalorNatSpeech*[18]; see **Real MEEG datasets** section) with a synthetic realistic *TRF₁* waveform, which was obtained by modelling the impulse response as a sum of Gaussian components with timing and amplitude corresponding to the P1–N1–P2 components matching empirical measurements based on low-frequency (1-8 Hz) EEG[19]. Specifically, the parameters used are reported as triplets i.e., ( amplitude [arbitrary units], latency [ms], width [ms] ): P1 (1, 40, 15), N1 (−2, 90, 20), P2 (1.5, 150, 25). The final synthetic EEG signal was obtained by adding noise ($y_n$) to the synthetic ground-truth, with noise consisting of randomly selected segments of low-frequency (1-8 Hz) EEG data (from the *LalorNatSpeech* dataset). To includes the spatial dimension, EEG noise was assigned to 64 scalp locations according to the standard BioSemi maps. Centring at the scalp location Fz, the ground-truth neural signal was assigned to the synthetic EEG with magnitude diminishing for sensors that are further away from Fz, with a rapid decay via the formula $y(ch) = y_1 \cdot \frac{1}{4^{\text{distance}_{Fz,ch}}} + y_n$. Thirty minutes of synthetic data was generated for each simulated participant, organised in ninety-second trials corresponding to different segments of the audiobook.

### SyntheticEEG-2

A similar procedure was followed for building a richer synthetic EEG dataset with two speech features from *LalorNatSpeech:* speech envelope ($x_1$) and word onsets ($x_2$). Each feature was convolved with a corresponding synthetic TRF waveform *TRF₁* and *TRF₂* to produce simulated neural signals $y_1$ and $y_2$. The final synthetic EEG ($y$) combines real EEG noise ($y_n$) with these simulated signals ($y_1 + y_2$). Six datasets (N=50 each) were created at mean SNRs of 0, −10, −20, −30, −40, and −50 dB, each with ±2.5 dB SNR variability (via the value α, which is randomly sampled from a uniformly distribution [-2.5,2.5] for each simulated participant) and adjustable $x_2$ encoding strength (β ∈ [0.02, 1]). This way, the multivariate synthetic EEG dataset can be used to study the ability of the encoding model to tease apart the neural representation of stimulus features of interest, as well as the sensitivity of the model to variations at the level of single participants, such as changes in SNR and β (which could be intuitively related with factors such as language proficiency in this simulation). Spatial maps were creating using the same spatial decay formula as for the univariate synthetic EEG, with $y_1$ centred in Fz and $y_2$ centred in Pz (**Fig. 2a**). Thirty minutes of synthetic data was generated for each of the fifty simulated participant in each SNR group, resulting in 150 hours of simulated data in total.

## Real MEEG datasets

The metrics presented in this study are tested on 34 datasets, comprising 31 EEG and 2 MEG experiments. An dataset recorded with functional Near-Infrared Spectroscopy (fNIRS) was also included, offering insights into the validity of the results on infrared-based hemodynamic measurements, which are substantially slower but, nonetheless, sensitive to neural auditory responses[67]. For simplicity, the full set of data used in this manuscript is referred to as *Real MEEG datasets,* albeit imprecisely as that also includes one fNIRS dataset.

Overall, the datasets include data from 573 participants, some of which were recorded across multiple sessions (e.g., GwilliamsSpeechMEG-1 and 2) or across different tasks that were stored as different datasets (e.g., Conversation Listening and Speaking), leading to 818 datapoints overall.

Tasks include listening, selective auditory attention, speaking, watching, and auditory imagery, with stimuli including speech (e.g., audiobooks, podcasts), vocoded speech, sung speech, music, sign language, and audio-visual material. The majority of the participants were young adults with no history of neurological impairment. In addition, data from children and infants (193 datapoints) is included (**Table 1**). In the majority of the datasets, experiments were carried out in a dark room. Participants were presented with auditory or visual stimuli while instructed to minimise motor movements. Auditory material included audiobooks, podcast recordings, MIDI melodies, sung speech. Video material included sign language recordings, infant-directed speech stories, and comedy show recordings. The full list of MEEG datasets, with links to the data and relevant papers, is available in **Supplementary Table 1**, and has also been shared at the link https://diliberg.net ('Open Data' section). All data was collected in accordance with the Declaration of Helsinki and was approved by local Ethics Committees. Each participant provided written informed consent. Further details on each datasets can be found at the links to the repository and relevant papers

## Data preprocessing

Analyses were conducted with MATLAB 2021a by using custom scripts developed starting from previous analysis scripts from my own work shared publicly as part of the CNSP initiative (Cognition and Natural Sensory Processing; https://cnspworkshop.net; see section Data and Code Availability for further details). All dataset were preprocessed with the same procedure, with the exception of *PodcastListeningfNIRS*, with filter cut-off frequencies and TRF time-lag window that were suitable for capturing the slow dynamics of hemodynamic responses.

EEG and MEG data were low-pass filtered at 8 Hz by means of zero-phase shift Butterworth filters with order 2 (by using the filtering functions in the CNSP resources) to include the Δ- and Θ-bands. EEG data were also high-pass filtered at 1 Hz to reduce noise and downsampled by an integer ratio to 32 or 25 Hz, depending on whether the original sampling frequency was a multiple of 2 (e.g., 512 Hz) or 10 (e.g., 500 Hz). Note that synthetic datasets were also preprocessed with the same procedure, with the only difference that the filtering operation was only applied to the EEG noise. The synthetic TRFs used to generate the signal, instead, were already incorporating that filtering operation, since they were built to resemble real TRFs from filtered data from the *LalorNatSpeech* dataset.

Channels with excessive noise were identified and interpolated via spherical interpolation, if they were three standard deviations away from the mean. EEG signals only were then re-referenced to the average of all channels. Continuous stimulus features were rescaled by dividing their values by the standard deviation calculated on each feature separately. Discrete stimulus features that were indicator variables with values one did not require a rescaling. All other discrete stimulus features were rescaled similar to the continuous stimulus features, but only by considering the non-zero values when calculating the standard deviation. Continuous MEEG signals were rescaled by dividing their values by the standard deviation after flattening the matrix across channel and time-samples dimensions, producing a rescaled X and Y that have comparable deviation, while preserving the relative magnitude across MEEG channels. All preprocessing operations were also repeated for the Δ- (0.5-4Hz) and Θ- (4-8 Hz) bands separately.

## Validation procedure

Validation of the ground-truth approximation procedure was carried out on univariate encoding models, by comparing the R-CPA and R-CPA-PA metrics with the conventional R-AVG and also with R-MAX, using both MEEG prediction correlations and signal detectability measurements. Multivariate models required a more elaborate procedure, which was aimed at determining the ability of the CPA procedure to retrieve model parameters of interest (i.e., β) while controlling for the sensitivity to SNR.

**Retrieving neural signatures at the single-participant**

The multivariate analysis in this study was designed to test the encoding model's ability to recover neural signatures of interest at single-participant level. TRF models were fit to map both the speech envelope and word onsets to the neural recordings. These two stimulus features are known to correlate with neural activity arising from at least two distinct spatial generators: the speech envelope typically predicts responses strongest over centro-temporal scalp regions, whereas word onset are also associated with responses over centro-parietal channels[4,28]. As such, this framework provides a suitable testbed for evaluating whether the encoding model can differentiate between two distinct neural sources.

CPA naturally returns multiple spatial components without requiring changes to the procedure. In contrast, the conventional evaluation approach required adaptation. Specifically, after fitting TRF models using both stimulus features simultaneously, predictions were generated using one feature at a time while partialling out the contribution of the other i.e., by setting the corresponding feature vector to zero. The R-MAX score derived from predictions based on word onsets (with envelope-related activity partialled out) is referred to as R-MAX-PARTIAL (**Fig. 2d**).

The resulting metrics R-MAX-PARTIAL (reflecting isolated word-onset responses), R-CPA-CC1, and R-CPA-CC2 (expected to reflect the envelope and word-onset activity respectively) were then correlated with β and SNR. An ideal evaluation metric would show a large correlation with the target variable β while remaining uncorrelated with fluctuations in SNR.

**Comparison with conventional CCA and overfitting assessment**

To better understand CPA, conventional CCA mappings are also fit in this study. This study adopts the CCA-1 and CCA-2 approaches as defined by de Cheveigné and colleagues[23]. CCA-1 performs a direct CCA mapping between $X_{lags}$ and Y, where the time-lags were applied to X but not Y. CCA-2 also applies the time-lags to Y, increasing the flexibility of the model. Specifically, CCA-1 rotates neural signal using spatial maps, thus constraining the transformation to signals that can be discriminated spatially. By adding the time-lags, CCA-2 enables separation of components by considering spatio-temporal transformations, increasing the rank and leading to a more powerful mapping with regard to variance explained. However, that flexibility comes at the cost of an increased risk of overfitting and a reduced interpretability.

This study includes an overfitting assessment, signal detectability was derived on the training and test folds for CPA, CCA-1, and CCA-2. A small degree of overfitting would lead to similar scores in the training and test folds, while a large overfitting would lead to high training scores and low test scores. Based on that rationale, an overfitting index was derived by subtracting signal detectability on the training and test folds at the individual participant level. The overfitting index was derived on a subset of the Real MEEG datasets, comprising 16 datasets with consistent task (listening), type of stimuli (continuous speech), and cohort (neurotypical adults). Specifically, the following datasets were included: *LalorNatSpeech, LalorRevSpeech,*

*AliceSpeech, MEGspeechGwilliams-1, MEGspeechGwilliams-2, PodcastListeningEEG, TrustSpeech, SparrKULee1, SparrKULee2, VocodedSpeech, ConversationListening, CocktailAttSwitch, AAD_KULeuven, EmotionSpeech, FDSpeech_L1, FDSpeech_L2*

## Statistical analysis

All statistical analyses directly comparing the metrics (R-AVG, R-MAX, R-CPA, R-CPA-PA) were performed using repeated measures ANOVA, with *F*-values reported as $F(df_{time}, df_{error})$ when the assumptions of normality and sphericity were met. Those assumptions were tested with Shapiro-Wilk's test and Mauspher's test respectively. Assumptions of normality for statistical tests were met unless otherwise stated. When the assumption of sphericity was not met, a Greenhouse-Geisser's correction was applied. Two-tailed paired Wilcoxon signed-rank tests were used for *post hoc* tests. Correction for multiple comparisons was applied where necessary via the false discovery rate (FDR) approach. The FDR-adjusted *p*-value was reported.

## Data availability

Data was converted to the CND data structure (Continuous-event Neural Data[68] - https://cnspworkshop.net), allowing to analysing all datasets with the same preprocessing and analysis scripts. The list of MEEG datasets, with links to the data and relevant papers, is available at https://cnspworkshop.net. **Supplementary Table 1** provides a complete list of datasets, each with the link to the original publication, authors list, link to the standardised CND version of the dataset, and link to the original dataset repository. Note that 17 datasets were available prior to this study, 11 datasets will be made available during the publication process of this manuscript, and the rest will be shared after an embargo period.

Analyses were conducted by using custom MATLAB scripts developed starting from publicly available scripts shared by the CNSP initiative (Cognition and Natural Sensory Processing; https://cnspworkshop.net). Such analysis scripts avail of external publicly available libraries: the mTRF-Toolbox (https://github.com/mickcrosse/mTRF-Toolbox)[8], EEGLAB[69]; and the NoiseTools library (http://audition.ens.fr/adc/NoiseTools)[49].

All scripts, including the new custom functions and data, can be downloaded from https://osf.io/c76p8/overview. Only the preprocessed data after downsampling was uploaded on the OSF dataset for minimising the size of the repository. The full datasets can be downloaded individually (see links in **Supplementary Table 1**) and preprocessed with the scripts provided. The scripts including future updates will be available on the CNSP resources GitHub repository (https://github.com/CNSP-Workshop/CNSP-resources/tree/main/CNSP).

# References

1. Holdgraf, C.R., Rieger, J.W., Micheli, C., Martin, S., Knight, R.T., and Theunissen, F.E. (2017). Encoding and Decoding Models in Cognitive Electrophysiology. Front Syst Neurosci *11*, 61. 10.3389/fnsys.2017.00061.
2. Crosse, M.J., Zuk, N.J., Di Liberto, G.M., Nidiffer, A.R., Molholm, S., and Lalor, E.C. (2021). Linear Modeling of Neurophysiological Responses to Speech and Other Continuous Stimuli: Methodological Considerations for Applied Research. Frontiers in neuroscience *15*, 705621-705621. 10.3389/fnins.2021.705621.
3. King, J.-R., and Gramfort, A. (2018). Encoding and decoding neuronal dynamics: Methodological framework to uncover the algorithms of cognition.
4. Broderick, M.P., Anderson, A.J., Di Liberto, G.M., Crosse, M.J., and Lalor, E.C. (2018). Electrophysiological Correlates of Semantic Dissimilarity Reflect the Comprehension of Natural, Narrative Speech. Current Biology. 10.1016/j.cub.2018.01.080.
5. Brodbeck, C., Hong, L.E., and Simon, J.Z. (2018). Rapid Transformation from Auditory to Linguistic Representations of Continuous Speech. Current Biology *28*, 3976-3983.e3975.
6. Leonard, M.K., Gwilliams, L., Sellers, K.K., Chung, J.E., Xu, D., Mischler, G., Mesgarani, N., Welkenhuysen, M., Dutta, B., and Chang, E.F. (2024). Large-scale single-neuron speech sound encoding across the depth of human cortex. Nature *626*, 593-602. 10.1038/s41586-023-06839-2.
7. de Cheveigné, A., and Simon, J.Z. (2008). Denoising based on spatial filtering. Journal of Neuroscience Methods *171*, 331-339. 10.1016/j.jneumeth.2008.03.015.
8. Crosse, M.J., Di Liberto, G.M., Bednar, A., and Lalor, E.C. (2016). The multivariate temporal response function (mTRF) toolbox: A MATLAB toolbox for relating neural signals to continuous stimuli. Frontiers in Human Neuroscience *10*. 10.3389/fnhum.2016.00604.
9. Lalor, E.C., Power, A.J., Reilly, R.B., and Foxe, J.J. (2009). Resolving Precise Temporal Processing Properties of the Auditory System Using Continuous Stimuli. Journal of Neurophysiology *102*, 349-359. 10.1152/jn.90896.2008.
10. Di Liberto, G.M., Barsotti, M., Vecchiato, G., Ambeck-Madsen, J., Del Vecchio, M., Avanzini, P., and Ascari, L. (2021). Robust anticipation of continuous steering actions from electroencephalographic data during simulated driving. Scientific Reports *11*, 23383. 10.1038/s41598-021-02750-w.
11. Nidiffer, A.R., Cao, C.Z., O'Sullivan, A., and Lalor, E.C. (2023). A representation of abstract linguistic categories in the visual system underlies successful lipreading. NeuroImage *282*, 120391. https://doi.org/10.1016/j.neuroimage.2023.120391.
12. Rogachev, A., and Sysoeva, O. (2024). Neural tracking of natural speech in children in relation to their receptive speech abilities. Cognitive Systems Research *86*, 101236.
13. Kalashnikova, M., Peter, V., Di Liberto, G.M., Lalor, E.C., and Burnham, D. (2018). Infant-directed speech facilitates seven-month-old infants' cortical tracking of speech. Scientific Reports *8*. 10.1038/s41598-018-32150-6.
14. Di Liberto, G.M., Pelofi, C., Bianco, R., Patel, P., Mehta, A.D., Herrero, J.L., De Cheveigné, A., Shamma, S., and Mesgarani, N. (2020). Cortical encoding of melodic expectations in human temporal cortex. Elife *9*, e51784.
15. Lalor, E.C., Pearlmutter, B.A., Reilly, R.B., McDarby, G., and Foxe, J.J. (2006). The VESPA: a method for the rapid estimation of a visual evoked potential. NeuroImage *32*, 1549-1561.
16. Brodbeck, C., Presacco, A., and Simon, J.Z. (2018). Neural source dynamics of brain responses to continuous stimuli: Speech processing from acoustics to comprehension. NeuroImage *172*, 162-174. https://doi.org/10.1016/j.neuroimage.2018.01.042.
17. Thornton, M., Mandic, D., and Reichenbach, T. (2022). Robust decoding of the speech envelope from EEG recordings through deep neural networks. Journal of neural engineering *19*, 046007.
18. Broderick, M., Anderson, A., Di Liberto, G., Crosse, M., and Lalor, E. (2018). Data from: electrophysiological correlates of semantic dissimilarity reflect the comprehension of natural, narrative speech. Dryad Digital Repository. Published online February 23, 2018.

19. Di Liberto, G.M., O'Sullivan, J.A., and Lalor, E.C. (2015). Low-frequency cortical entrainment to speech reflects phoneme-level processing. Current Biology *25*. 10.1016/j.cub.2015.08.030.
20. O'Sullivan, J.A., Power, A.J., Mesgarani, N., Rajaram, S., Foxe, J.J., Shinn-Cunningham, B.G., Slaney, M., Shamma, S.A., and Lalor, E.C. (2014). Attentional Selection in a Cocktail Party Environment Can Be Decoded from Single-Trial EEG. Cerebral Cortex, bht355-bht355.
21. Vanthornhout, J., Decruy, L., Wouters, J., Simon, J.Z., and Francart, T. (2018). Speech Intelligibility Predicted from Neural Entrainment of the Speech Envelope. Journal of the Association for Research in Otolaryngology *19*, 181-191. 10.1007/s10162-018-0654-z.
22. Ding, N., Chatterjee, M., and Simon, J.Z. (2014). Robust cortical entrainment to the speech envelope relies on the spectro-temporal fine structure. NeuroImage *88*, 41-46.
23. de Cheveigné, A., Wong, D.E., Di Liberto, G.M., Hjortkjær, J., Slaney, M., and Lalor, E. (2018). Decoding the auditory brain with canonical component analysis. NeuroImage *172*, 206-216. 10.1016/j.neuroimage.2018.01.033.
24. Di Liberto, G.M., Wong, D., Melnik, G.A., and de Cheveigne, A. (2019). Low-frequency cortical responses to natural speech reflect probabilistic phonotactics. NeuroImage *196*, 237-247. 10.1016/j.neuroimage.2019.04.037.
25. Geirnaert, S., Vandecappelle, S., Alickovic, E., de Cheveigné, A., Lalor, E., Meyer, B.T., Miran, S., Francart, T., and Bertrand, A. (2021). Neuro-Steered Hearing Devices: Decoding Auditory Attention From the Brain.
26. Hannah, J., and Di Liberto, G.M. (2026). Trust Modulates Speech Entrainment: Enhanced Cortical Tracking for Low Trust Speakers. bioRxiv, 2026.2003. 2011.711118.
27. Ip, E.Y., Akkaya, A., Winchester, M.M., Bishop, S.J., Cowan, B.R., and Di Liberto, G.M. (2025). Exploring the impact of social relevance on the cortical tracking of speech: viability and temporal response characterisation. bioRxiv, 2025.2009. 2023.674728.
28. Chalehchaleh, A., Winchester, M.M., and Di Liberto, G. (2024). Robust assessment of the cortical encoding of word-level expectations using the temporal response function. Journal of Neural Engineering, 2024-2004. https://doi.org/10.1088/1741-2552/ada30a.
29. Jessen, S., Obleser, J., and Tune, S. (2021). Neural tracking in infants – An analytical tool for multisensory social processing in development. Developmental Cognitive Neuroscience *52*, 101034. https://doi.org/10.1016/j.dcn.2021.101034.
30. Herbst, S.K., Fiedler, L., and Obleser, J. (2018). Tracking Temporal Hazard in the Human Electroencephalogram Using a Forward Encoding Model. eneuro *5*, ENEURO.0017-0018.2018. 10.1523/ENEURO.0017-18.2018.
31. Di Liberto, G.M., Peter, V., Kalashnikova, M., Goswami, U., Burnham, D., and Lalor, E.C. (2018). Atypical cortical entrainment to speech in the right hemisphere underpins phonemic deficits in dyslexia. NeuroImage *NIMG-17-29*, 70-79. 10.1016/J.NEUROIMAGE.2018.03.072.
32. De Cheveigné, A., Wong, D.D.E., Di Liberto, G.M., Hjortkjær, J., Slaney, M., and Lalor, E. (2018). Decoding the auditory brain with canonical component analysis. NeuroImage *172*, 206-216.
33. Jessen, S., Fiedler, L., Münte, T.F., and Obleser, J. (2019). Quantifying the individual auditory and visual brain response in 7-month-old infants watching a brief cartoon movie. NeuroImage *202*, 116060-116060. 10.1016/j.neuroimage.2019.116060.
34. Di Liberto, G.M., Attaheri, A., Cantisani, G., Reilly, R.B., Ní Choisdealbha, Á., Rocha, S., Brusini, P., and Goswami, U. (2023). Emergence of the cortical encoding of phonetic features in the first year of life. Nature Communications *14*, 7789.
35. Attaheri, A., Choisdealbha, Á.N., Di Liberto, G.M., Rocha, S., Brusini, P., Mead, N., Olawole-Scott, H., Boutris, P., Gibbon, S., and Williams, I. (2022). Delta-and theta-band cortical tracking and phase-amplitude coupling to sung speech by infants. NeuroImage *247*, 118698.
36. Agmon, G., Jaeger, M., Tsarfaty, R., Bleichner, M.G., and Zion Golumbic, E. (2023). "Um…, it's really difficult to… um… speak fluently": Neural tracking of spontaneous speech. Neurobiology of Language *4*, 435-454.
37. Solanki, V.J. (2017). Brains in dialogue: investigating accommodation in live conversational speech for both speech and EEG data. (University of Glasgow).

38. Van de Ryck, I., Heintz, N., Rotaru, I., Geirnaert, S., Bertrand, A., and Francart, T. (2026). EEG-based Decoding of Auditory Attention to Conversations with Turn-taking Speakers. Hearing Research, 109539.
39. Orf, M., Tune, S., Hannemann, R., and Obleser, J. (2025). Speech, gait, and brain dynamics during natural conversation in motion. bioRxiv, 2025.2012.2023.696153. 10.64898/2025.12.23.696153.
40. King, J.-R., Charton, F., Lopez-Paz, D., and Oquab, M. (2020). Back-to-back regression: Disentangling the influence of correlated factors from multivariate observations. NeuroImage *220*, 117028.
41. de Cheveigné, A., Di Liberto, G.M., Arzounian, D., Wong, D., Hjortkjaer, J., Fuglsang, S.A., and Parra, L.C. (2018). Multiway Canonical Correlation Analysis of Brain Signals. bioRxiv, 344960-344960. 10.1101/344960.
42. Accou, B., Vanthornhout, J., hamme, H.V., and Francart, T. (2023). Decoding of the speech envelope from EEG using the VLAAI deep neural network. Scientific Reports *13*, 812. 10.1038/s41598-022-27332-2.
43. Broderick, M., Di Liberto, G., Anderson, A., Rofes, A., and Lalor, E. (2020). Dissociable electrophysiological measures of natural language processing reveal differences in speech comprehension strategy in healthy ageing. bioRxiv, 2020.2004.2017.046201-042020.046204.046217.046201. 10.1101/2020.04.17.046201.
44. Mischler, G., Li, Y.A., Bickel, S., Mehta, A.D., and Mesgarani, N. (2024). Contextual Feature Extraction Hierarchies Converge in Large Language Models and the Brain. arXiv preprint arXiv:2401.17671.
45. Broderick, M.P., Di Liberto, G.M., Anderson, A.J., Rofes, A., and Lalor, E.C. (2021). Dissociable electrophysiological measures of natural language processing reveal differences in speech comprehension strategy in healthy ageing. Scientific Reports *11*, 4963. 10.1038/s41598-021-84597-9.
46. Di Liberto, G.M., Nie, J., Yeaton, J., Khalighinejad, B., Shamma, S.A., and Mesgarani, N. (2021). Neural representation of linguistic feature hierarchy reflects second-language proficiency. NeuroImage *227*, 117586-117586. 10.1016/j.neuroimage.2020.117586.
47. Di Liberto, G.M., Pelofi, C., Shamma, S., and de Cheveigné, A. (2020). Musical expertise enhances the cortical tracking of the acoustic envelope during naturalistic music listening. Acoustical Science and Technology *41*.
48. Cohen, S.S., and Parra, L.C. (2016). Memorable Audiovisual Narratives Synchronize Sensory and Supramodal Neural Responses. eneuro *3*, ENEURO.0203-0216.2016. 10.1523/eneuro.0203-16.2016.
49. de Cheveigné, A., Di Liberto, G.M., Arzounian, D., Wong, D.D.E., Hjortkjær, J., Fuglsang, S., and Parra, L.C. (2019). Multiway canonical correlation analysis of brain data. NeuroImage *186*, 728-740. 10.1016/J.NEUROIMAGE.2018.11.026.
50. Obleser, J., and Kayser, C. (2019). Neural Entrainment and Attentional Selection in the Listening Brain. Trends in Cognitive Sciences. Elsevier Ltd.
51. Klimovich-Gray, A., Di Liberto, G., Amoruso, L., Barrena, A., Agirre, E., and Molinaro, N. (2023). Increased top-down semantic processing in natural speech linked to better reading in dyslexia. NeuroImage *273*, 120072. https://doi.org/10.1016/j.neuroimage.2023.120072.
52. Attaheri, A., Ní Choisdealbha Á, Di Liberto, G.M., Rocha, S., Brusini, P., Mead, N., Olawole-Scott, H., Boutris, P., Gibbon, S., Williams, I., et al. (2022). Delta- and theta-band cortical tracking and phase-amplitude coupling to sung speech by infants. Neuroimage *247*, 118698. 10.1016/j.neuroimage.2021.118698.
53. Verschueren, E., Somers, B., and Francart, T. (2019). Neural envelope tracking as a measure of speech understanding in cochlear implant users. Hearing Research *373*, 23-31. https://doi.org/10.1016/j.heares.2018.12.004.
54. Somers, B., Verschueren, E., and Francart, T. (2018). Neural tracking of the speech envelope in cochlear implant users. Journal of Neural Engineering *16*, 16003-16003. 10.1088/1741-2552/aae6b9.
55. Chen, Y.-P., Schmidt, F., Keitel, A., Rösch, S., Hauswald, A., and Weisz, N. (2023). Speech intelligibility changes the temporal evolution of neural speech tracking. NeuroImage *268*, 119894. https://doi.org/10.1016/j.neuroimage.2023.119894.

56. Van Hirtum, T., Somers, B., Dieudonné, B., Verschueren, E., Wouters, J., and Francart, T. (2023). Neural envelope tracking predicts speech intelligibility and hearing aid benefit in children with hearing loss. Hearing Research *439*, 108893. https://doi.org/10.1016/j.heares.2023.108893.
57. Van Hirtum, T., Somers, B., Verschueren, E., Dieudonné, B., and Francart, T. (2023). Delta-band neural envelope tracking predicts speech intelligibility in noise in preschoolers. Hearing Research *434*, 108785. https://doi.org/10.1016/j.heares.2023.108785.
58. Carta, S., Aličković, E., Zaar, J., Valdés, A.L., and Di Liberto, G.M. (2025). Simultaneous cortical tracking of competing speech streams during attention switching. bioRxiv, 2025.2007. 2002.662762.
59. Haufe, S., Meinecke, F., Görgen, K., Dähne, S., Haynes, J.-D., Blankertz, B., and Bießmann, F. (2014). On the interpretation of weight vectors of linear models in multivariate neuroimaging. NeuroImage *87*, 96-110. https://doi.org/10.1016/j.neuroimage.2013.10.067.
60. Di Liberto, G.M., Marion, G., and Shamma, S.A. (2021). The Music of Silence: Part II: Music Listening Induces Imagery Responses. The Journal of Neuroscience *41*, 7449. 10.1523/JNEUROSCI.0184-21.2021.
61. Marion, G., Di Liberto, G.M., and Shamma, S.A. (2021). The Music of Silence. Part I: Responses to Musical Imagery Accurately Encode Melodic Expectations and Acoustics. Journal of Neuroscience.
62. Varma, S., and Simon, R. (2006). Bias in error estimation when using cross-validation for model selection. BMC Bioinformatics *7*, 91. 10.1186/1471-2105-7-91.
63. Wong, D.D.E., Fuglsang, S.A., Hjortkjaer, J., Ceolini, E., Slaney, M., and De Cheveigne, A. (2018). A Comparison of Regularization Methods in Forward and Backward Models for Auditory Attention Decoding. Frontiers in Neuroscience *12*, 531-531. 10.3389/FNINS.2018.00531.
64. Di Liberto, G.M., and Lalor, E.C. (2017). Indexing cortical entrainment to natural speech at the phonemic level: Methodological considerations for applied research. Hearing Research *348*, 70-77. 10.1016/j.heares.2017.02.015.
65. Brookshire, G., Lu, J., Nusbaum, H.C., Goldin-Meadow, S., and Casasanto, D. (2017). Visual cortex entrains to sign language. Proceedings of the National Academy of Sciences of the United States of America *114*, 6352-6357. 10.1073/pnas.1620350114.
66. Ding, N., and Simon, J.Z. (2012). Neural coding of continuous speech in auditory cortex during monaural and dichotic listening. Journal of Neurophysiology *107*, 78-89. 10.1152/jn.00297.2011.
67. Wilroth, J., Silva, N.S., Tafakkor, A., de Avo Mesquita, B., Ip, E.Y.J., Lau, B., Hannah, J., and Di Liberto, G.M. (2026). Investigating neural speech processing with functional near infrared spectroscopy: considerations for temporal response functions. bioRxiv, 2026.2003.2020.713212. 10.64898/2026.03.20.713212.
68. Di Liberto, G.M., Nidiffer, A., Crosse, M.J., Zuk, N., Haro, S., Cantisani, G., Winchester, M.M., Igoe, A., McCrann, R., and Chandra, S. (2024). A standardised open science framework for sharing and re-analysing neural data acquired to continuous stimuli. Neurons, Behavior, Data analysis, and Theory, 1-25.
69. Delorme, A., and Makeig, S. (2004). EEGLAB: an open source toolbox for analysis of single-trial EEG dynamics including independent component analysis. J Neurosci Methods *134*, 9-21. 10.1016/j.jneumeth.2003.10.009.
70. Brennan, J.R. (2018). EEG Datasets for Naturalistic Listening to "Alice in Wonderland" (Version 1).
71. Gwilliams, L., Flick, G., Marantz, A., Pylkkänen, L., Poeppel, D., and King, J.-R. (2023). Introducing MEG-MASC a high-quality magneto-encephalography dataset for evaluating natural speech processing. Scientific Data *10*, 862. 10.1038/s41597-023-02752-5.
72. Piazza, G., Carta, S., Ip, E., Pérez-Navarro, J., Kalashnikova, M., Martin, C.D., and Di Liberto, G.M. (2024). Are you talking to me? How the choice of speech register impacts listeners' hierarchical encoding of speech. bioRxiv, 2024-2009.
73. Piazza, G., Carta, S., Ip, E.Y.J., Pérez-Navarro, J., Kalashnikova, M., Martin, C.D., and Di Liberto, G.M. (2025). Are you talking to me? How the choice of speech register impacts listeners' hierarchical encoding of speech.
74. Bollens, L., Accou, B., Van hamme, H., and Francart, T. (2023). SparrKULee: A Speech-evoked Auditory Response Repository of the KU Leuven, containing EEG of 85 participants. V3 ed. KU Leuven RDR.

75. De Palma, I.C., Lopez, L.S., and Lopez-Valdes, A. (2023). Effects of spectral and temporal modulation degradation on intelligibility and cortical tracking of speech signals. pp. 5192-5196.
76. Di Liberto, G.M., Goswami, U., Attaheri, A., Choisdealbha Á, N., Rocha, S., Mead, N., Olawole-Scott, H., and Grey, C. (2023). Data and code from "Emergence of the cortical encoding of phonetic features in the first year of life".  OSF https://osf.io/mdnwg.
77. Das, N., Francart, T., and Bertrand, A. (2019). Auditory Attention Detection Dataset KULeuven (Version 2.0).  Zenodo.
78. Cantisani, G., Shamma, S., and Di Liberto, G.M. (2024). Neural signatures of musical and linguistic interactions during natural song listening.
79. Di Liberto, G.M., Pelofi, C., Bianco, R., Patel, P., Mehta, A.D., Herrero, J.L., De Cheveigné, A., Shamma, S.A., and Mesgarani, N. (2021). Cortical encoding of melodic expectations in human temporal cortex.
80. Di Liberto, G.M., Marion, G., and Shamma, S.A. (2021). Data from The music of silence: Part ii: Music listening induces imagery responses. . Dryad.  Dryad.

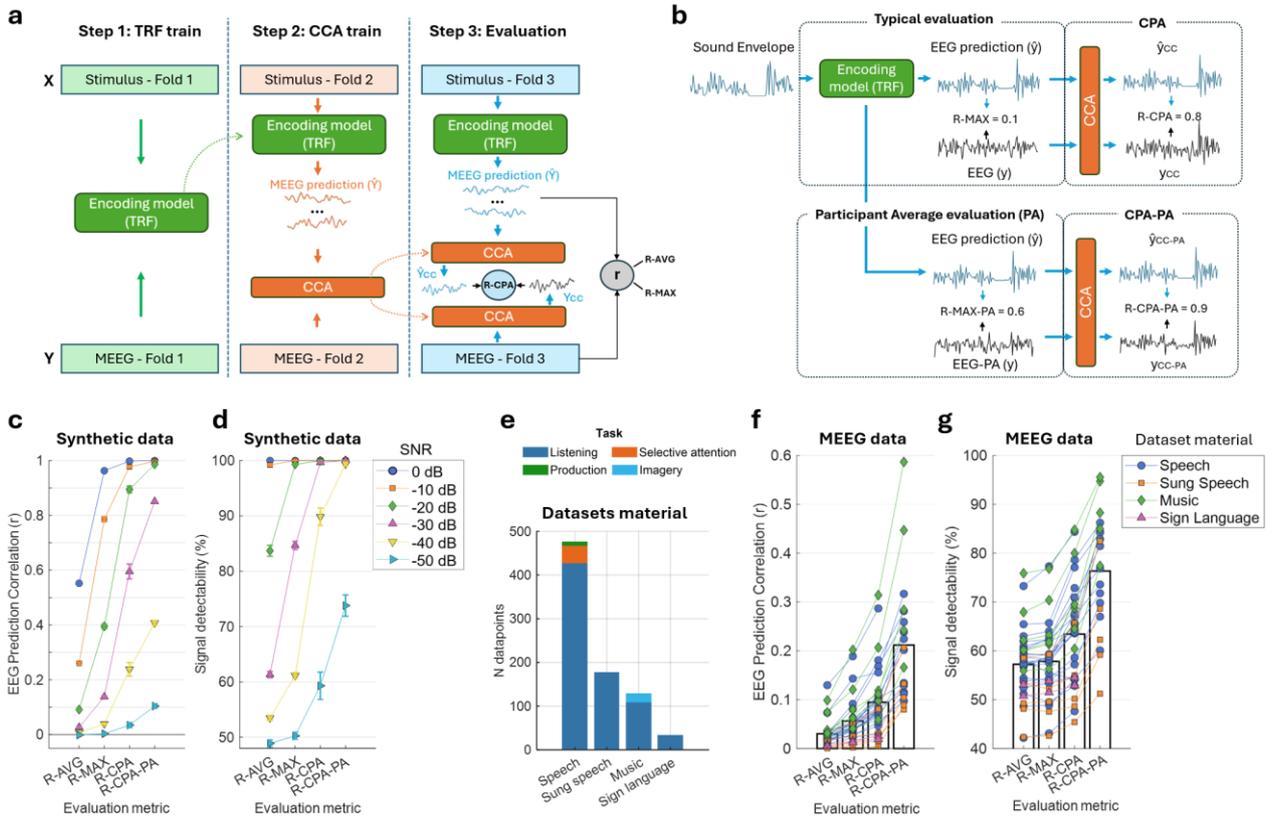

**Fig. 1 | Evaluation method overview and performance on univariate encoding models. a**, Canonical Prediction Alignment (CPA) workflow. Data are partitioned into three folds. The encoding model is trained on Fold 1 and used to generate MEEG predictions on the other folds. Fold 2 is used to fit a CCA mapping between predicted and measured MEEG signals, which is then applied to Fold 3 to compute the correlation of the first canonical component (R-CPA). This procedure is repeated under cross-validation. **b**, Illustration of conventional evaluation R-MAX, CPA, participant-average evaluation (PA), and CPA-PA on the SyntheticEEG-1 dataset at an SNR of -30dB. Conventional metrics yield low correlations due to EEG noise, even at the best channel (R-MAX = 0.1). CPA projects predictions and acquired neural signal to a shared canonical space, attenuating the task-irrelevant noise in the neural signal, approximating the underlying ground-truth signal. When neural signals can be temporally aligned across participants, participant averaging further refines the ground-truth approximation, yielding R-CPA-PA. **c,** Evaluation of univariate envelope encoding models on synthetic EEG across SNRs (N=30 participants per SNR). Conventional metrics (R-AVG, R-MAX) are compared with R-CPA and R-CPA-PA; Mean ± SE are shown. **d,** Signal detectability quantified as classification accuracy (%) in identifying whether a 5-s EEG prediction segment corresponds to the correct synthetic EEG segment or a random one. **e,** Overview of datasets and tasks. Thirty-four datasets were analysed (31 EEG, 2 MEG, 1 fNIRS), comprising 818 participant-level datapoints. **f,g,** MEEG prediction correlations and signal detectability for univariate envelope encoding models fitted to low-frequency (1-8 Hz) signals across 34 real MEEG datasets. Each dot represents the participant average for a given dataset.

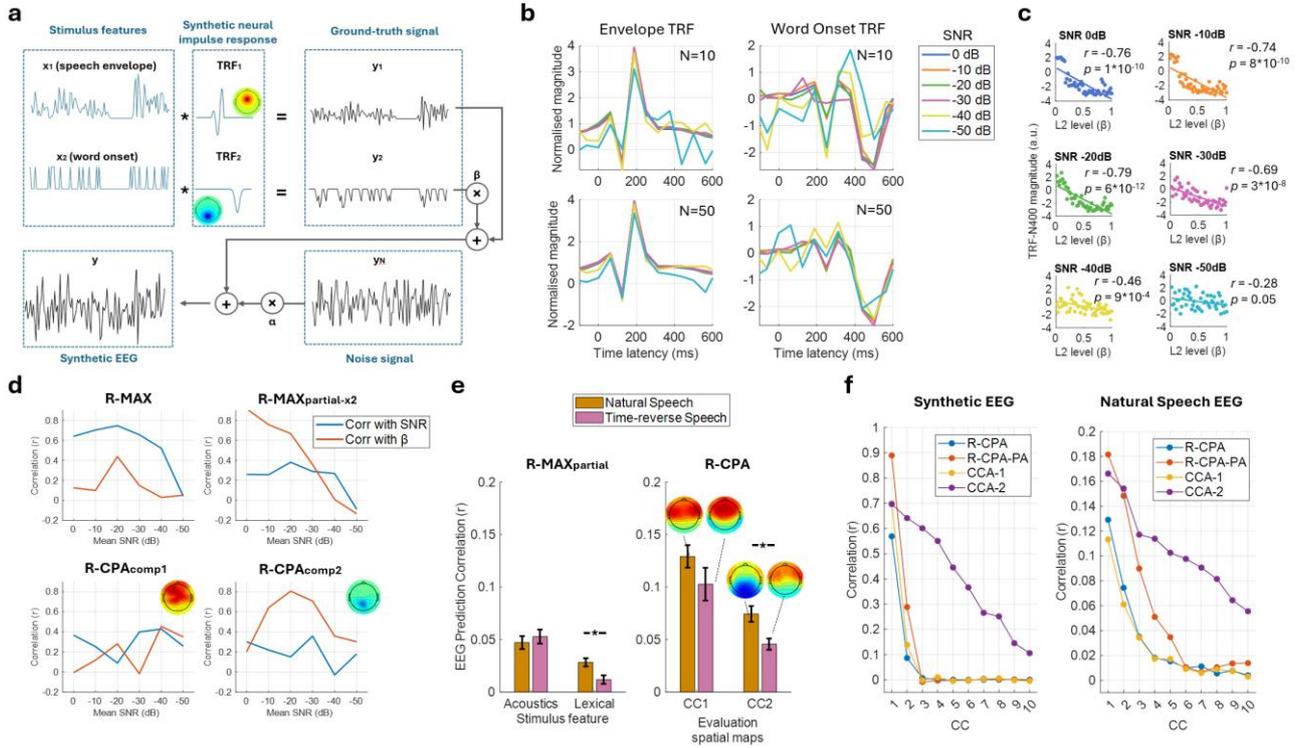

**Fig. 2 | Robust retrieval of neural variables of interests with multivariate encoding models. a**, Synthetic EEG generation pipeline. Speech features from *LalorNatSpeech* ($x_1$: envelope; $x_2$: word onsets) were convolved with synthetic impulse responses to generate simulated neural signals ($y_1$, $y_2$), which were combined with real EEG noise ($y_n$) to form the final signal ($y$). Data were generated for six groups (N=50 each) at mean SNRs of 0 to -50dB, with ±2.5dB variability (α) and variable $x_2$ encoding strength (β). **b**, Multivariate encoding models fitted on synthetic data using speech envelope and word-onset features. Group-averaged TRFs at Fz (envelope) and Pz (word onset) recover the ground-truth impulse responses across all SNRs. Estimates degrade when fewer participants are averaged (N=10). **c**, Trough magnitude at 450 ms (TRF-N400) correlates significantly with β at all SNRs, with reduced correlation values at lower SNRs, indicating successful recovery of $x_2$ encoding strength across noise levels. **d**. Sensitivity of evaluation metrics to β and SNR. R-CPA (first two canonical components, CCs) is compared with an EEG prediction metric isolating $x_2$ (R-MAX$_{partial-x2}$). The ideal metric correlates strongly with β but minimally with SNR. **e**, Retrieval of neural variables from real EEG during natural (*LalorNatSpeech*) and time-reversed (*LalorRevSpeech*) speech listening. Acoustic and lexical (word onset) features were modelled, with R-MAX$_{partial}$ isolating the contributions of acoustic (envelope, $x_1$) and lexical (word onset, $x_2$) features. R-CPA is reported for the first two CCs. Robust acoustic signatures is observed in both datasets, whereas lexical encoding is present only for natural speech. **f**, R-CPA and R-CPA-PA scores across CCs. In synthetic data with two neural sources at -30dB, stimulus-relevant information is captured primarily by the first two CCs. In real MEEG data, 3-5 CCs contain stimulus-relevant information, consistent with stimulus-MEEG CCA using stimulus-only lags (CCA-1). Applying lags to both stimulus and neural signals (CCA-2) yields a broader dispersion of information across CCs.

| Name | Participants | | Task | | | Technology | | Country |
|---|---|---|---|---|---|---|---|---|
| | Age | N | Stimulus type | Modality | Type | Device | Channels | |
| LalorNatSpeech[4,18,19] | Adults | 19 | Speech | A | Listening | EEG BioSemi Active-Two | 128 | Ireland |
| LalorRevSpeech[4,18,19] | Adults | 10 | Speech | A | Listening | EEG BioSemi Active-Two | 128 | Ireland |
| AliceSpeech[70] | Adults | 20 | Speech | A | Listening | EEG actiCap, Brain Products GmbH | 60 | USA |
| GwilliamsSpeechMEG-1[71] | Adults | 11 | Speech | A | Listening | MEG 208 axial-gradiometer | 208 | USA |
| GwilliamsSpeechMEG-2[71] | Adults | 10 | Speech | A | Listening | MEG 208 axial-gradiometer | 208 | USA |
| PodcastListening[27,*] | Adults | 20 | Speech | A | Listening | EEG BioSemi Active-Two | 64 | Ireland |
| PodcastListening fNIRS[67,*] | Adults | 8 | Speech | A | Listening co-presence | fNIRS Artinis | 16 | Netherlands |
| TrustSpeech[26,*] | Adults | 20 | Speech | A | Listening | EEG BioSemi Active-Three | 64 | Ireland |
| EmotionSpeech[27,*] | Adults | 27 | Speech | A | Listening | EEG BioSemi Active-Two | 64 | Ireland |
| FDSpeech L1[72,73] | Adults | 19 | Speech | A | Listening | EEG BioSemi Active-Two | 64 | Ireland |
| FDSpeech L2[72,73] | Adults | 21 | Speech | A | Listening | EEG Brain Products GmbH | 64 | Basque |
| SparrKULee1[74] | Adults | 77 | Speech | A | Listening | EEG BioSemi Active-Two | 64 | Belgium |
| SparrKULee2[74] | Adults | 56 | Speech | A | Listening | EEG BioSemi Active-Two | 64 | Belgium |
| VocodedSpeech[75,*] | Adults | 13 | Vocoded Speech | A | Listening | EEG BioSemi Active-Two | 64 | Ireland |
| ChildStories_Sysoeva[12] | 3-8yo children | 52 | Speech | A | Listening | EEG Brain Products actiCHamp | 32 | Russia |
| BabyRhythmCambridge Adults[34,76] | Adults | 17 | Sung speech | AV | Listening | EEG GES 300 amplifier, Geodesic | 64 | UK |
| BabyRhythmCambridge 4mo[34,76] | 4mo infants | 47 | Sung speech | AV | Listening | EEG GES 300 amplifier, Geodesic | 64 | UK |
| BabyRhythmCambridge 7mo[34,76] | 7mo infants | 47 | Sung speech | AV | Listening | EEG GES 300 amplifier, Geodesic | 64 | UK |
| BabyRhythmCambridge 11mo[34,76] | 11mo infants | 47 | Sung speech | AV | Listening | EEG GES 300 amplifier, Geodesic | 64 | UK |
| ConversationListening* | Adults | 10 | Speech | A | Listening | EEG mBrainTrain Smarting | 24 | Ireland |
| ConversationSpeaking* | Adults | 20 | No stimulus | A | Speaking | EEG mBrainTrain Smarting | 24 | Ireland |
| CocktailAttSwitch[58,*] | Adults | 24 | Speech | A | Listening selective attention | EEG BioSemi Active-Two | 64 | Ireland |
| AAD KULeuven[77] | Adults | 16 | Speech | A | Listening selective attention | EEG BioSemi Active-Two | 64 | Belgium |
| CantisaniSpeech[78] | Adults | 20 | Speech | A | Listening | EEG BioSemi Active-Two | 64 | France |
| CantisaniMelody[78] | Adults | 20 | Music | A | Listening | EEG BioSemi Active-Two | 64 | France |
| CantisaniSong[78] | Adults | 20 | Sung speech | A | Listening | EEG BioSemi Active-Two | 64 | France |
| diliBach[14,79] | Adults | 20 | Music | A | Listening | EEG BioSemi Active-Two | 64 | France |
| MelodySwitch* | Adults | 17 | Music | A | Listening | EEG BioSemi Active-Two | 64 | Ireland |
| PolyphonicBach* | Adults | 30 | Music | A | Listening | EEG mBrainTrain Smarting | 24 | Ireland |
| MusicImagery Listening[60,61,80] | Adults | 21 | Music | A | Listening | EEG BioSemi Active-Two | 64 | France |
| MusicImagery Imagination[60,61,80] | Adults | 21 | Metronome | - | Auditory imagery | EEG BioSemi Active-Two | 64 | France |
| StandupComedy* | Adults | 14 | Speech/Video | AV | Listening | EEG BioSemi Active-Two | 64 | Ireland |
| SignLanguageSigners | Adults | 14 | Sign Language | V | Watching | EEG BioSemi Active-Two | 64 | France |
| SignLanguageNonsigners | Adults | 20 | Sign Language | V | Watching | EEG BioSemi Active-Two | 64 | France |

**Table 1 | List of datasets** including the essential information on each real dataset analysed in this study. Please see Supplementary Table 1 for additional details, including links to the datasets, relevant papers, stimulus material, secondary tasks. The star symbol next to the dataset title indicates new datasets that were not publicly available previously (note that some of them may have an embargo period). The shading in the table groups datasets (e.g., same experiment with different cohorts or recording device).

**a**

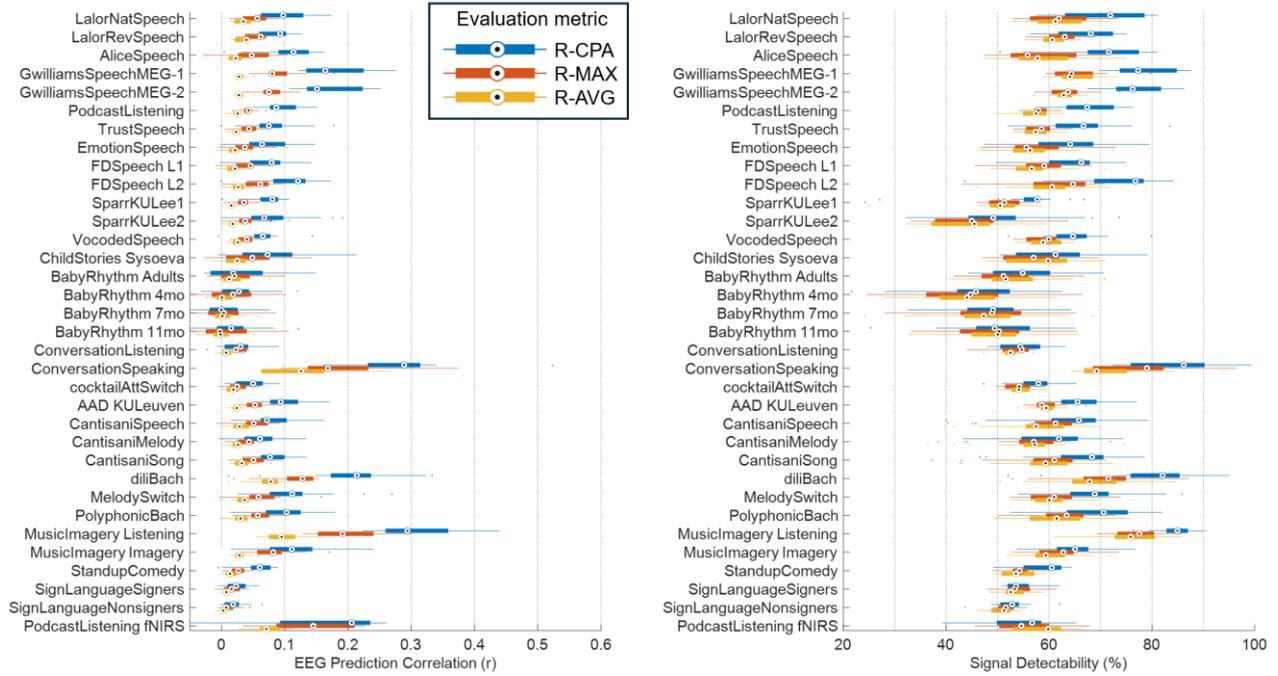

**b**

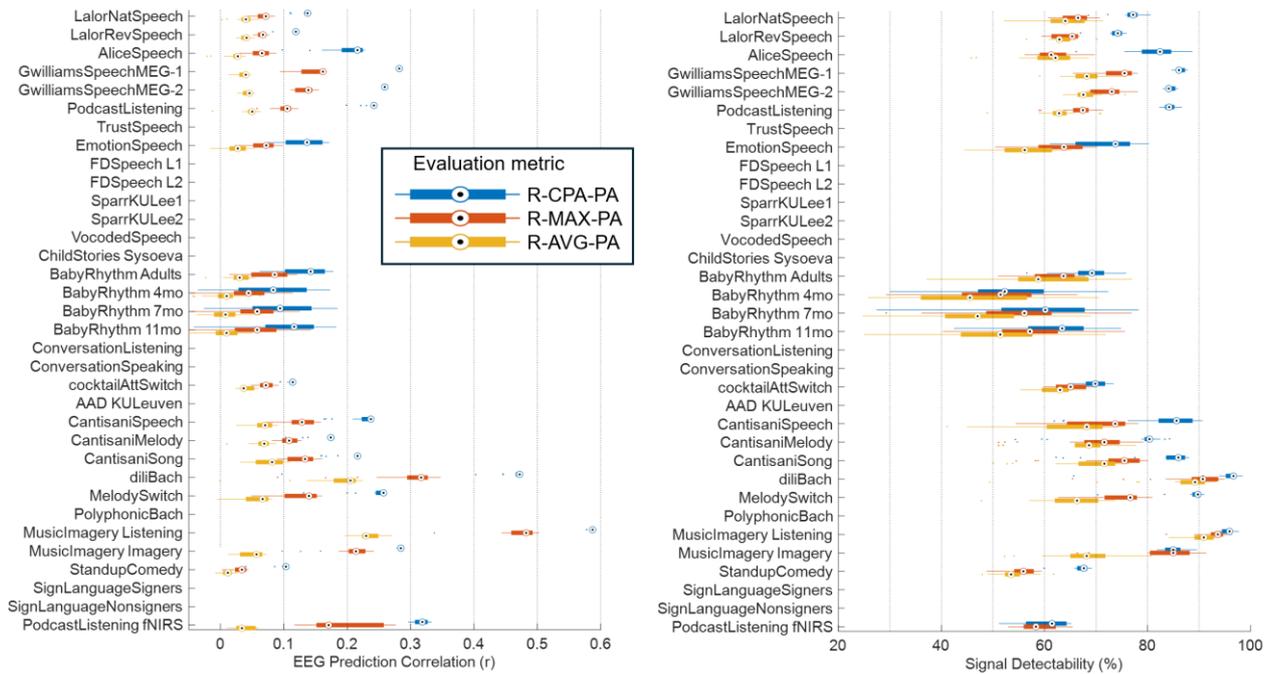

**Supplementary Fig. 1 | Evaluation of univariate encoding models fit to real datasets.** This figure extends **Fig. 1f,g**, presenting the result distributions for each dataset. Low-frequency MEEG (1-8 Hz) and fNIRS (0.01-0.7 Hz) prediction correlations (left) and signal detectability (right) are compared for the three evaluation metrics. Boxplots represent the 25[th] and 75[th] percentiles of the result distribution across participants, with the circle capturing the median value and the whiskers indicating the extreme non-outlier values. Outliers are shows are separate dots when identified. **a**, Comparing R-CPA with R-MAX and R-AVG. **b**, A similar comparison is shown when participant averaging (PA) evaluation is applied (R-CPA-PA, R-MAX-PA, R-AVG-PA). This analysis is possible for datasets where the neural data across participants have a precise temporal correspondence (e.g., participants are presented with the same identical stimuli).

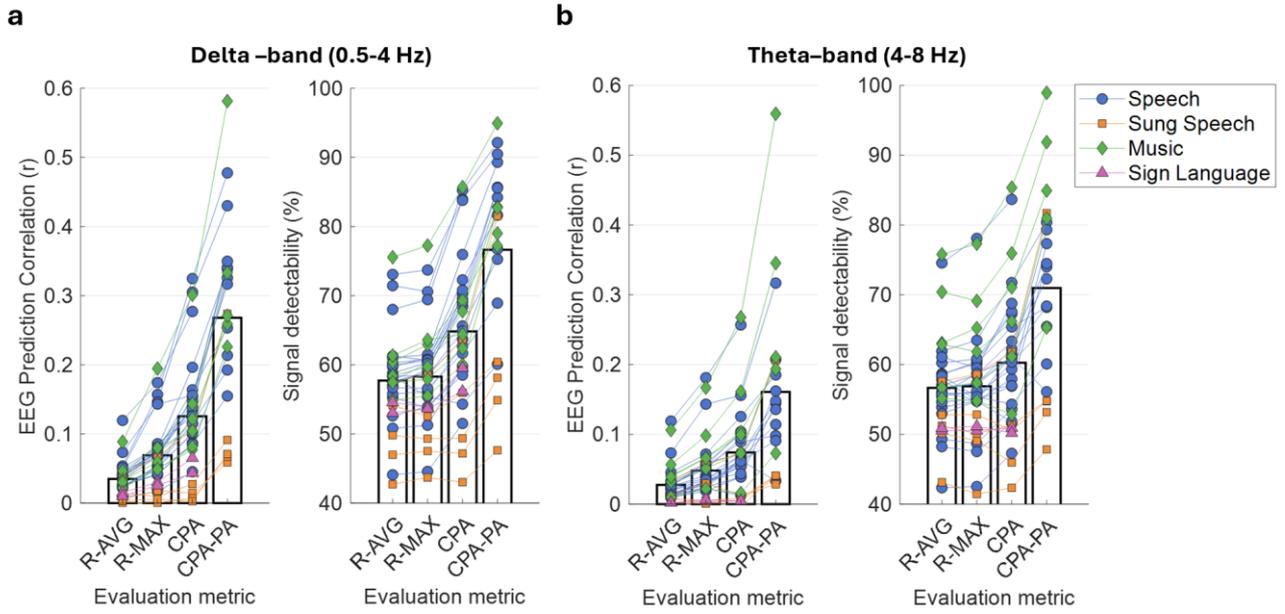

**Supplementary Fig. 2 | Univariate encoding model performance across neural rates.** Correlation and signal detectability distributions are shown for each real dataset, extending Fig. 1f,g. **a**, Delta-band (0.5–4 Hz) MEEG. **b**, Theta-band (4–8 Hz). Each dot denotes the participant average per dataset.

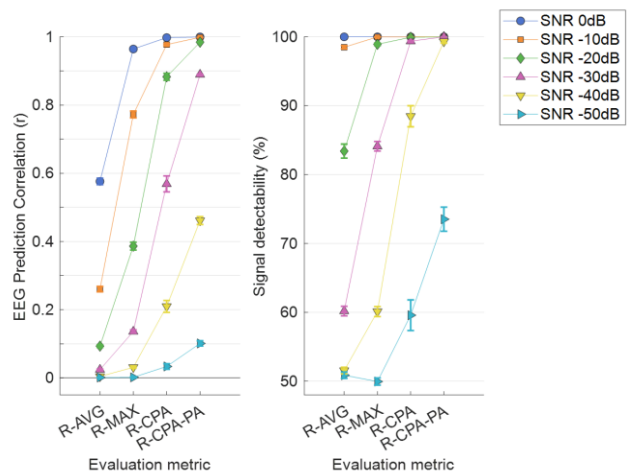

**Supplementary Fig. 3 | Multivariate encoding model evaluation on synthetic data.** Equivalent to Fig. 1c, shown for SyntheticEEG-2. Each SNR condition includes N=50 synthetic participants with individual SNRs uniformly distributed within ±2.5 dB of the mean.